\documentclass[twocolumn,aps,nofootinbib]{revtex4-1}

\usepackage{graphicx}
\usepackage{epstopdf}
\usepackage{latexsym}
\usepackage{amsmath}
\usepackage{amssymb}
\usepackage{color}
\usepackage{mathrsfs}

\usepackage[center]{subfigure}

\begin{document}

 \newcommand{\bq}{\begin{equation}}
 \newcommand{\eq}{\end{equation}}
 \newcommand{\bqn}{\begin{eqnarray}}
 \newcommand{\eqn}{\end{eqnarray}}
 \newcommand{\nb}{\nonumber}
 \newcommand{\lb}{\label}
\newcommand{\PRL}{Phys. Rev. Lett.}
\newcommand{\PL}{Phys. Lett.}
\newcommand{\PR}{Phys. Rev.}
\newcommand{\CQG}{Class. Quantum Grav.}

\title{Cosmic evolution of dark energy in a generalized Rastall gravity}

\author{Kai Lin $^{a, b}$}\email{lk314159@hotmail.com}
\author{Wei-Liang Qian$^{c,b,d}$}\email{wlqian@usp.br}

\affiliation {a) Hubei Subsurface Multi-scale Imaging Key Laboratory, Institute of Geophysics and Geomatics, China University of Geosciences, Wuhan 430074, Hubei, China}
\affiliation{b) Escola de Engenharia de Lorena, Universidade de S\~ao Paulo, 12602-810, Lorena, SP, Brazil}
\affiliation{c) Center for Gravitation and Cosmology, College of Physical Science and Technology, Yangzhou University, Yangzhou 225009, China}
\affiliation{d) Faculdade de Engenharia de Guaratinguet\'a, Universidade Estadual Paulista, 12516-410, Guaratinguet\'a, SP, Brazil}
%

\date{Mar. 24, 2020}

\begin{abstract}
In this work, we propose a scheme for cosmic evolution in a generalized Rastall gravity.
In our approach, the role of dark energy is taken by the non-conserved sector of the stress energy-momentum tensor.
The resultant cosmic evolution is found to naturally consists of three stages, namely, radiation dominated, ordinary matter dominated, as well as dark energy and dark matter dominated eras.
Furthermore, for the present model, it is demonstrated that the eventual fate of the Universe is mostly insensitive to the initial conditions, in contrast to the standard $\Lambda$CDM model.
In particular, the solution displays the properties of a dynamic attractor, which is reminiscent of quintessence and k-essence models.
Subsequently, the cosmic coincidence problem is averted.
The amount of deviation from a conserved stress energy-momentum tensor is shown to be more remarkable during the period when the dark energy evolves more rapidly.
On the other hand, the conservation law is largely restored for the infinite past and future.
The implications of the present approach are addressed.

\end{abstract}


\maketitle

\section{Introduction}
\renewcommand{\theequation}{1.\arabic{equation}} \setcounter{equation}{0}

As an alternative to general relativity, Rastall gravity is characterized by the modified conservation law of the stress energy-momentum tensor (SET) in curved spacetime~\cite{agr-rastall-01}.
The theory implies intriguing novelty in various aspects regarding black hole physics~\cite{agr-rastall-02, agr-rastall-03, agr-rastall-04, agr-rastall-05, agr-rastall-06, agr-rastall-09, agr-rastall-11, agr-rastall-15, agr-rastall-16, agr-rastall-21, agr-rastall-22} and cosmology~\cite{agr-rastall-cosmo-01, agr-rastall-cosmo-02, agr-rastall-cosmo-03, agr-rastall-cosmo-04, agr-rastall-cosmo-05, agr-rastall-cosmo-06, agr-rastall-cosmo-07, agr-rastall-cosmo-08, agr-rastall-cosmo-09, agr-rastall-cosmo-10} as it has been explored recently by many authors.
In particular, the rudimentary feature of Rastall gravity, in a natural manner, supplies an alternative implementation for the dark energy.

The potential limit of general relativity has been systematically investigated on the largest scale against various observational data, namely, the supernova, large scale structure, and the cosmic microwave background (CMB) measurements.
Among others, one of the most significant findings is the apparent accelerating expansion of the Universe, and subsequently, the dark energy scenario has become the most accepted premise regarding a satisfactory account for the experimental data.
Moreover, it is deduced that the Universe at the present day is mostly composed of dark energy and dark matter. 
Subsequently, the physical properties, as well as the cosmic evolution of dark energy, become an increasingly active area in cosmology~\cite{agr-cosmo-review-01, agr-cosmo-review-02, agr-cosmo-review-03, agr-cosmo-review-04, agr-cosmo-review-05}, due to its immediate connection with our understanding of the fundamental nature of the Universe.

Although the standard $\Lambda$ Cold Dark Matter ($\Lambda$CDM) model supplies a reasonable account for the observed properties of the cosmos, it also confronts several challenges such as cosmic coincidence problem and fine tuning problem.
In this regard, alternative approaches are primarily carried out by modifying Einstein's field equations, which can be further divided into two distinct categories. 
The first type of model focuses on the properties of the matter field, which gives rise to dynamical dark energy models.
In the literature, efforts along this train of thought consist of quintessence~\cite{agr-cosmo-de-quintessence-01}, tachyon~\cite{agr-cosmo-de-tachyon-01}, k-essence~\cite{agr-cosmo-de-k-essence-01}, phantom~\cite{agr-cosmo-de-phantom-01}, Chaplygin gas~\cite{agr-cosmo-de-chaplygin-01}, holographic dark energy~\cite{agr-cosmo-de-holo-01, agr-cosmo-de-holo-02, agr-cosmo-de-holo-04}, agegraphic dark energy~\cite{agr-cosmo-de-agegraphic-01, agr-cosmo-de-agegraphic-02}, among others.
The second type of approach, on the other hand, is motivated by generalizing the geometry in Einstein's general relativity.
Such attempts include $f(R)$~\cite{agr-modified-gravity-fR-review-01}, $f(T)$~\cite{agr-modified-gravity-fT-review-01}, $f(R, T)$ theory~\cite{agr-modified-gravity-fRT-01}, Brans-Dicke theory~\cite{agr-modified-gravity-BD-01}, Gauss-Bonnet theory~\cite{agr-modified-gravity-GB-01}, Lovelock~\cite{agr-modified-gravity-Lovelock-01}, and Horava-Lifshitz theories~\cite{agr-modified-gravity-horava-01, agr-modified-gravity-horava-02, agr-modified-gravity-horava-lw-01, agr-modified-gravity-horava-lw-05}.

In general relativity, the SET is minimally coupled to the geometry.
Consider a matter field that possesses a classical continuous symmetry, and a conserved current is implied according to the Noether theorem.
However, as an infinitesimal symmetry transformation is made local, the action is no longer invariant, but rather it gives rise to a contribution associated with the Noether current.
The above spacetime dependent transformation is a well-known procedure of introducing a gauge field into the theory. 
Here, the metric is playing the role of the gauge field for a diffeomorphism invariance, and the latter is related to the translation symmetry of the original theory.
Subsequently, the Hilbert energy-momentum tensor, defined by the variation of the action of the matter field with respect to the metric, is conserved.
In this context, it has been argued that the Rastall gravity can be viewed such that the curvature-matter coupling is implemented by a non-minimal fashion~\cite{agr-rastall-23}.
Therefore, the theory might be classified into the second category of modified gravity. 

From a physical viewpoint, both the gravitationally induced particle production~\cite{agr-cosmo-bb-04, agr-cosmo-bb-05, agr-cosmo-bb-06} and quantum effects in curved spacetime~\cite{book-qft-Bertlemann} might be associated with the violations of the usual conservation law of the SET.
This particularly meaningful as it is understood that the conservation of SET does not lead to particle production~\cite{book-qft-curved-space-Birrell-Davies}.
From the viewpoint of relativistic kinetic theory, there is one more apparent mechanism even if the particle number is conserved, namely, the kinetic diffusive process.
As it was pointed out in Ref.~\cite{agr-modified-gravity-FP-01}, the SET of the matter field is not conserved, as the evolution of the matter field is governed by the Fokker-Planck equation.
Moreover, it can be shown that the divergence of the SET equals to a conserved four-current. 
In the study of cosmology, the above physical scenarios are relevant and evidently lead to important implications.
In particular, the non-conserved part of the SET might give rise to the dark energy which, subsequently, is responsible for the present accelerating expansion of the Universe~\cite{agr-rastall-cosmo-02, agr-rastall-cosmo-03, agr-rastall-cosmo-04, agr-rastall-cosmo-09, agr-rastall-cosmo-05, agr-rastall-cosmo-10}. 
In Ref.~\cite{agr-rastall-cosmo-10}, the authors studied the accelerating expansion of the Universe by employing a generalized Rastall theory.
In particular,  a non-minimal coupling between the geometry and a pressureless matter field is shown to lead the transition from the matter-dominated era to the accelerating expansion.
The cosmic evolution is also investigated for homogeneous and isotropic flat Friedmann–Lema\^itre–Robertson–Walker (FLRW) metric in Ref.~\cite{agr-rastall-cosmo-11}.
The model is shown to be equivalent to the particle creation mechanism in Einstein gravity in the framework of non-equilibrium thermodynamics.

The present study involves such an attempt to construct a reasonable scheme for cosmic evolution in a generalized Rastall gravity.
In our model, the dark energy is implemented so that it is closely related to the violation of SET.
The amount of violation is found to be more significant during the period when the contribution of dark energy increases and raises to its present value.
It eventually becomes insignificant, as it is naturally dictated by the equations of motion.
The resultant cosmic evolution experiences three stages, namely, radiation dominated, ordinary matter dominated, as well as dark energy and dark matter dominated eras.
We also show that the eventual fate of the Universe is insensitive to the initial conditions, owing to the dynamical attractor behavior of the solution.

The rest of the paper is organized as follows.
In the following section, we briefly discuss the generalized Rastall gravity utilized in the present study.
The equations of motion of the cosmic evolution are derived in section III.
Numerical results are presented in section IV.
Concluding remarks are given in the last section.

\section{Generalized Rastall gravity}
\renewcommand{\theequation}{2.\arabic{equation}} \setcounter{equation}{0}

In Refs.~\cite{agr-rastall-21, agr-rastall-22}, based on the original idea by Rastall~\cite{agr-rastall-01}, we proposed a generalized formulation of the Rastall theory.
To be specific, the equation of the gravitational field equation and that of the SET read
\bqn
\lb{RastallEquation1}
R_{\mu\nu}-\frac{1}{2}g_{\mu\nu}R&=&\kappa\left(T_{\mu\nu}-{\mathscr A}_{\mu\nu}\right) ,\nb\\
\nabla_\mu {T^{\mu}}_{\nu}&=&\nabla_\mu {{\mathscr A}^{\mu}}_{\nu},
\eqn
where $\kappa=8\pi G$.
We also impose a physical requirement that the effect of ${{\mathscr A}^{\mu}}_{\nu}$ and its derivatives must vanish in flat spacetime.
In fact, it can be shown that the above formulation is rather general so that several modified gravity theories could be viewed as its special cases~\cite{agr-rastall-21}.

As for the purpose of the present study, we consider a specific case, namely,
\bqn
\lb{RastallEquation2}
{\mathscr A}_{\mu\nu}=\lambda g_{\mu\nu} {\mathscr H} ,
\eqn
where ${\mathscr H}$ vanishes when $R=0$.
On the other hand, as a scalar, ${\mathscr H}$ can be a function of the Ricci scalar $R$, $T\equiv g^{\mu\nu}T_{\mu\nu}$ and other constants. 
By substituting the form of ${\mathscr A}_{\mu\nu}$ into Eq.~(\ref{RastallEquation1}), we have
\bqn
\lb{RastallEquation3}
R_{\mu\nu}-\frac{1}{2}g_{\mu\nu}R&=&\kappa\left(T_{\mu\nu}-\lambda g_{\mu\nu}{\mathscr H}\right),\nb\\
\nabla_\mu {T^{\mu}}_{\nu}&=&\lambda \nabla_\nu {\mathscr H} .
\eqn
For algebraic convenience, one defines 
\bqn
\lb{taumunu}
\tau_{\mu\nu}=T_{\mu\nu}-\lambda g_{\mu\nu}{\mathscr H} .
\eqn
Therefore, Eq.~\eqref{RastallEquation3} can be rewritten in essentially the same form as in general relativity
\bqn
\lb{RastallEquation4}
R_{\mu\nu}-\frac{1}{2}g_{\mu\nu}R&=&\kappa\tau_{\mu\nu} ,\nb\\
\nabla_\mu {\tau^{\mu}}_{\nu}&=&0 .
\eqn
Although it is mathematically similar, usually, it is not physically appropriate to interpret $\tau^{\mu\nu}$ as the SET of the matter field~\cite{agr-rastall-18}.
If one contracts both sides of the gravitational field equation, it gives
\bqn
\lb{RastallEquation5}
R+\kappa T=4\kappa\lambda {\mathscr H}.
\eqn

Owing to the reasons to be discussed below, we choose
\bqn
\lb{H1}
{\mathscr H}=\frac{R\left(R+\kappa T\right)}{4\kappa\lambda\left(4\kappa\Theta-\kappa T\right)} ,
\eqn
where $\Theta$ is to be determined shortly.
We note that, in the vacuum, both factors on the numerator vanish as $R\to 0$.
In order that ${\mathscr H}$ is a well-defined quantity, one requires that the denominator of Eq.~\eqref{H1} being regular even when $R\to 0$.

By substituting ${\mathscr H}$ into Eq.~(\ref{RastallEquation5}), one finds a quadratic algebraic equation, which implies the following two solutions for $R$:
\bqn
\lb{H2}
R&=&-\kappa T,\nb\\
\text{or}~~~~~R&=&\kappa\left(4\Theta-T\right) .
\eqn
The first solution is not physically relevant, because here, we will investigate the scenario where $R$ remains finite even when the matter field $T_{\mu\nu}$ vanishes.
This is precisely the case where dark energy plays a significant role in cosmic evolution.
Therefore, we will only explore the implication of the second solution. 
For the present model, this implies that $\Theta$ is nonvanishing, while $T_{\mu\nu}$ vanishes.
This, in turn, ensures that the denominator of Eq.~\eqref{H1} will be regular in our approach.

By substituting it back into field equations, one finds
\bqn
\lb{GeneralRastall1}
R_{\mu\nu}-\frac{1}{2}g_{\mu\nu}R+\kappa g_{\mu\nu}\Theta&=&\kappa T_{\mu\nu}, \nb\\
\nabla_\mu {T^{\mu}}_{\nu}&=& \nabla_\nu \Theta .
\eqn

Before proceeding further, we pause to give a few comments regarding Eq.~\eqref{GeneralRastall1}.
First, if one assumes $\Theta\equiv\Lambda_{\text{eff}}/\kappa$ where $\Lambda_{\text{eff}}$ is a constant, the above equations become identical to those of the standard $\Lambda$CDM model.
Therefore, it seems rather appealing to identify the physical content of $\Theta$ with the cosmological constant.
Although, in the present model, as further discussed below, its temporal dependence plays an essential role.
In Ref.~\cite{agr-rastall-22}, it is demonstrated that an (anti-)de Sitter solution can be effectively found in Rastall gravity where the spacetime is asymptotically flat.
It is achieved by taking ${\mathscr H}={\mathscr H}(R)$ and $T_{\mu\nu}=0$.
In other words, the above solution again confirms the previous findings that a metric in asymptotically flat Rastall gravity naturally gives rise to that in general relativity with a cosmological constant. 
Moreover, according to the second equation of Eq.~\eqref{GeneralRastall1}, $\Theta$ measures the violation of the SET.
Indeed, from the viewpoint of the Rastall gravity, all different types of matter fields are described by $T_{\mu\nu}$, as a result, the observation of dark energy merely reflects, to what degree, the SET of the matter field deviates from a conserved current.
It is also worth mentioning that Eq.~\eqref{GeneralRastall1} is very similar to those obtained from different theories where the conservation of the SET is partly breaking (for instance, see Refs.~\cite{agr-modified-gravity-two-measure-03,agr-modified-gravity-two-measure-04} and related discussions in the last section).

In the following section, we proceed to derive the equations for cosmic evolution and investigate their solutions.
Accordingly, we will treat $\Theta$ as a variable, and solve its temperoal dependence.

\section{Cosmic evolution in generalized Rastall gravity}
\renewcommand{\theequation}{3.\arabic{equation}} \setcounter{equation}{0}

The equations for cosmic expansion can be formulated by employing the co-moving coordinates, in terms of which the SET of the matter field is given by
\bqn
\lb{Tensor1}
{T^{\mu}}_{\nu}=
\left(
  \begin{array}{cccc}
    -\rho & 0 & 0 & 0\\
    0 & P & 0 & 0\\
    0 & 0 & P & 0\\
    0 & 0 & 0 & P\\
  \end{array}
\right) .
\eqn

According to the discussions in the previous section, we denote $\Theta\equiv\rho_{\mathrm{de}}$, the energy density of the dark energy.
It is noted, by using Eq.~\eqref{taumunu} and the solution Eq.~\eqref{H2}, it is straightforward to show that the tensor ${\tau^{\mu}}_{\nu}$ reads
\bqn
\lb{Tensor2}
{\tau^{\mu}}_{\nu}=
\left(
  \begin{array}{cccc}
    -\rho-\rho_{\mathrm{de}} & 0 & 0 & 0\\
    0 & P+P_{\mathrm{de}} & 0 & 0\\
    0 & 0 & P+P_{\mathrm{de}} & 0\\
    0 & 0 & 0 & P+P_{\mathrm{de}}\\
  \end{array}
\right) ,\nb\\
\eqn
where $P_{\mathrm{de}}=-\rho_{\mathrm{de}}$ is recognized as the pressure of dark energy.
In other words, although $\Theta$ is not a constant, the equation of state of the dark energy still satisfies a simple form, namely, $w_{\mathrm{de}}=\frac{P_{\mathrm{de}}}{\rho_{\mathrm{de}}}=-1$, which is in agreement with the observed results.
Furthermore, the the cosmological principle implies that $\rho=\rho(t)$, $P=P(t)$, $\rho_{\mathrm{de}}=\rho_{\mathrm{de}}(t)$, and $P_{\mathrm{de}}=P_{\mathrm{de}}(t)$ are functions independent on spatial coordinates.

We proceed to derive the equations of motion in terms of the FLRW metric
\bqn
\lb{Metric1}
ds^2=-dt^2+a(t)^2\left[\frac{dr^2}{1-kr^2}+r^2\left(d\theta^2+\sin(\theta)^2d\varphi^2\right)\right] ,\nb\\
\eqn
where $k$ represents the curvature density of the Universe. 
Therefore, the $(0,0)$ and $(1,1)$ components of gravitational field equation in Eqs.~\eqref{GeneralRastall1} can be rewritten as
\bqn
\lb{CosmologyEquations1}
\left(\frac{\dot{a}}{a}\right)^2+\frac{k}{a^2}&=&\frac{8\pi G}{3}\left(\rho+\rho_{\mathrm{de}}\right),\nb\\
\frac{2\ddot{a}}{a}+\left(\frac{\dot{a}}{a}\right)^2+\frac{k}{a^2}&=&-8\pi G\left(P-\rho_{\mathrm{de}}\right),
\eqn
while the equation regarding the SET gives
\bqn
\lb{CosmologyEquations2}
\dot{\rho}+\dot{\rho}_{\mathrm{de}}=-\left(\rho+P\right)\frac{\dot{a}}{a} .
\eqn
We note only two of the above three equations are independent.

We consider the matter content of the Universe consists of radiation, ordinary matter, dark matter, and dark energy.
Radiation, ordinary matter, and dark matter are assumed to be independent between one another.
They satisfy the standard equations of states, namely, $P_{\mathrm{r}}=\frac{1}{3}\rho_{\mathrm{r}}$ and $P_{\mathrm{m}}=P_{\mathrm{dm}}=0$.
Therefore, the total pressure and density of the matter fields are given by
\bqn
\lb{Matter1}
\rho&=&\rho_{\mathrm{r}}+\rho_{\mathrm{m}}+\rho_{\mathrm{dm}},\nb\\
P&=&P_{\mathrm{r}}+P_{\mathrm{m}}+P_{\mathrm{dm}}=P_{\mathrm{r}}=\frac{1}{3}\rho_{\mathrm{r}} .
\eqn

As independent fluid components, we further assume that radiation and ordinary matter satisfy, respectively, an equation regarding the conservation of its SET, namely,
\bqn
\lb{CosmologyEquations3}
\dot{\rho}_{\mathrm{m}}+\rho_{\mathrm{m}}\frac{\dot{a}}{a}&=&0,\nb\\
\dot{\rho}_{\mathrm{r}}+\left(\rho_{\mathrm{r}}+\frac{1}{3}\rho_{\mathrm{r}}\right)\frac{\dot{a}}{a}&=&0 ,
\eqn
For the dark matter, however, the corresponding equation is constrained by Eq.~\eqref{CosmologyEquations2}.
It is not difficult to show that the resultant equation reads
\bqn
\lb{CosmologyEquations4}
\dot{\rho}_{\mathrm{dm}}+\dot{\rho}_{\mathrm{de}}+\rho_{\mathrm{dm}}\frac{\dot{a}}{a}=0 .
\eqn
Now, there is only one free variable left, and for the last equation, we impose a rather simple scenario:
\bqn
\lb{CosmologyEquations5}
\rho_{\mathrm{de}}=\beta\rho_{\mathrm{dm}} ,
\eqn
which can be viewed as to effectively incorporate a specific type of interaction between the dark energy and dark matter.
We note that this is in tune with the fact that Eq.~\eqref{CosmologyEquations4} also demonstrates that dark matter and dark energy are related.
Otherwise, in comparison with Eq.~\eqref{CosmologyEquations3}, the second term on the l.h.s. of Eq.~\eqref{CosmologyEquations4} would have not been present.

Eqs.~\eqref{CosmologyEquations3}, \eqref{CosmologyEquations4} and \eqref{CosmologyEquations5} can be solved analytically to give
\bqn
\lb{rho1}
{\rho}_{\mathrm{dm}}&=&{\rho}_{\mathrm{dm}0}\left(\frac{a_0}{a}\right)^{\frac{3}{1+\beta}},\nb\\
{\rho}_{\mathrm{m}}&=&{\rho}_{\mathrm{m}0}\left(\frac{a_0}{a}\right)^{3},\nb\\
{\rho}_{\mathrm{r}}&=&{\rho}_{\mathrm{r}0}\left(\frac{a_0}{a}\right)^{4} .
\eqn
Here, the radiation and ordinary matter evolve as in standard $\Lambda$CDM model.
Also, the evolution of the dark energy accompanies that of dark matter, which reads
\bqn
\lb{rho2}
{\rho}_{\mathrm{de}}&=&\beta\rho_{\mathrm{dm}}=\beta{\rho}_{\mathrm{dm}0}\left(\frac{a_0}{a}\right)^{\frac{3}{1+\beta}} .
\eqn
Here, the index $0$ indicates the values at present.

One can also rewrite the field equation similar to the Friedman equation.
By introducing the Hubble parameter $H\equiv\frac{\dot{a}}{a}$ and
and the spatial curvature density
\bqn
\lb{rho3}
{\rho}_{\mathrm{k}}&\equiv&-\frac{3k}{8\pi G a^2}={\rho}_{\mathrm{k}0}\left(\frac{a_0}{a}\right)^{2} .
\eqn
one finds
\bqn
\lb{CosmologyEquation6}
\Omega_{\mathrm{r}}+\Omega_{\mathrm{m}}+\Omega_{\mathrm{dm}}+\Omega_{\mathrm{de}}+\Omega_{\mathrm{k}}=1,
\eqn
where the $\Omega_i=\frac{8\pi G\rho_i}{3H^2}$ with $i=\mathrm{r, m, dm, de, k}$ indicating the density parameters of radiation, ordinary matter, dark matter, dark energy, and spatial curvature respectively.

The deceleration parameter $q\equiv-\frac{\ddot{a}a}{\dot{a}^2}$ is found to be
\bqn
\lb{CosmologyEquation7}
q=\Omega_{\mathrm{r}}-\Omega_{\mathrm{de}}+\frac{\Omega_{\mathrm{m}}+\Omega_{\mathrm{dm}}}{2} .
\eqn

\section{Numerical results}
\renewcommand{\theequation}{4.\arabic{equation}} \setcounter{equation}{0}

In the section, we present the numerical results in Figs.~\ref{Fig1_ev_Rastall}-\ref{Fig3_violation_Rastall}.
We first determine the constants of the integration regarding equations of the cosmic evolution by the values of the measurements to date~\cite{agr-cosmo-data-planck-01}.
To be specific, we choose $\Omega_{\mathrm{dm}0}=0.27$, $\Omega_{\mathrm{de}0}=0.68$, $\Omega_{\mathrm{m}0}=0.05$.
Also, we assume a spatially flat Universe by considering $k=0$.
Moreover, the redshift $z=1100$, where the energy density of ordinary matter exceeds that of the radiation, is also taken as an input~\cite{book-cosmology-Gregory-Zeilik}.
Subsequently, for the proposed model, the parameter $\beta$ is found to be $2.52$, which will be used in the remainder of this paper.
The calculations are then carried out for the generalized Rastall theory, which are compared against those from the standard $\Lambda$CDM model.
The corresponding results obtained by adopting the above parameters are shown in Figs.~\ref{Fig1_ev_Rastall}-\ref{Fig2_q_LCDM} by the solid curves in different colors.

On top of the above solution, the initial conditions of the relevant equations are arbitrarily perturbed at an instant with a sufficiently large redshift.
In other words, due to the perturbations, the constants of integration will no longer remain unchanged, and the calculations are performed for a system of five equations, namely, Eqs.~\eqref{CosmologyEquations1}, \eqref{CosmologyEquations3}, \eqref{CosmologyEquations4}, and \eqref{CosmologyEquations5} for five variables $a$, $\rho_{\mathrm{de}}$, $\rho_{\mathrm{dm}}$, $\rho_{\mathrm{r}}$, and $\rho_{\mathrm{m}}$ for given $\beta$.
Subsequently, we investigate how the evolution of the composition of the Universe, and in particular, the density parameters at present day $a/a_0=1$, depends on different initial conditions.
The latter are presented in dashed and dotted curves in Figs.~\ref{Fig1_ev_Rastall}, \ref{Fig1_ev_LCDM}, \ref{Fig2_q_Rastall}, and \ref{Fig2_q_LCDM} for both models.

As expected, from Fig.~\ref{Fig1_ev_Rastall}, the results show that the cosmic evolution consists of three stages, namely, the radiation dominated, ordinary matter dominated, as well as dark energy and dark matter dominated eras.
Also, it can be clearly inferred that the eventual fate of the Universe, calculated by the present model, is insensitive to the initial conditions.
To be specific, the density parameters for the dark energy and dark matter all converge to the given values, irrelevant to specific initial conditions.
Meanwhile, during the evolution, the compositions of the radiation and ordinary matter reflet the details of the perturbed initial conditions. 
This point becomes particularly evident as one compares the above results against those of the standard $\Lambda$CDM model shown in Fig.~\ref{Fig1_ev_LCDM}.
In the $\Lambda$DCM model, the density parameters at present $a/a_0=1$ are dictated largely by the initial conditions, as shown by the zoomed-in plot of Fig.~\ref{Fig1_ev_LCDM}.
We note that the present findings are in agreement with other approaches~\cite{agr-cosmo-de-holo-03, agr-cosmo-de-holo-04}, which incorporate the interaction between the dark energy and dark matter.
The difference for the present model is that, in the framework of Rastall theory, the dark energy degree of freedom appears naturally from the deviation from the conservation law of the SET.

\begin{figure}[tbp]
\centering
\includegraphics[width=1\columnwidth]{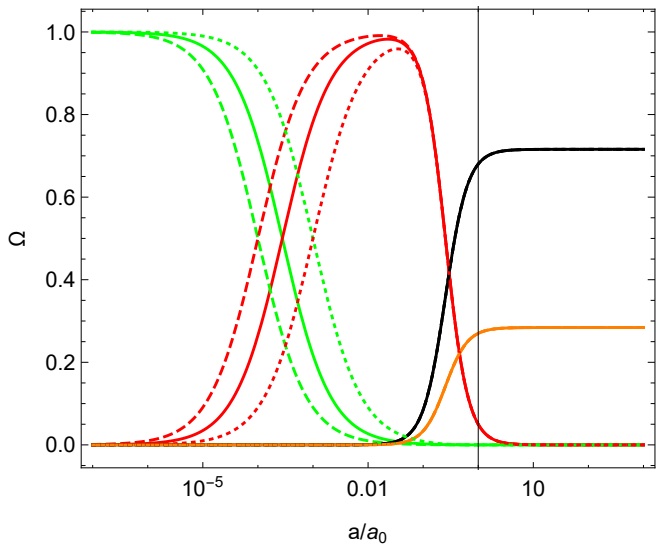}
\caption{The calcualted cosmic evolutions of dimensionless density parameters as functions of $a/a_0$.
The present day $a/a_0=1$ is indicated by a vertical black solid line.
The quantities $\Omega_{\mathrm{de}}, \Omega_{\mathrm{dm}}, \Omega_{\mathrm{m}}$, and $\Omega_{\mathrm{r}}$ are represented by black, orange, red,  and green curves.
The calculations are carried out for different parameters in generalized Rastall gravity.
The cosmic evolution evaluated by using the specific initial conditions which reproduces the measurements is presented by solid curves.
Those obtained by using different perturbed initial conditions are indicated by dashed and dotted curves.
}
\lb{Fig1_ev_Rastall}
\end{figure}

\begin{figure}[tbp]
\centering
\includegraphics[width=1\columnwidth]{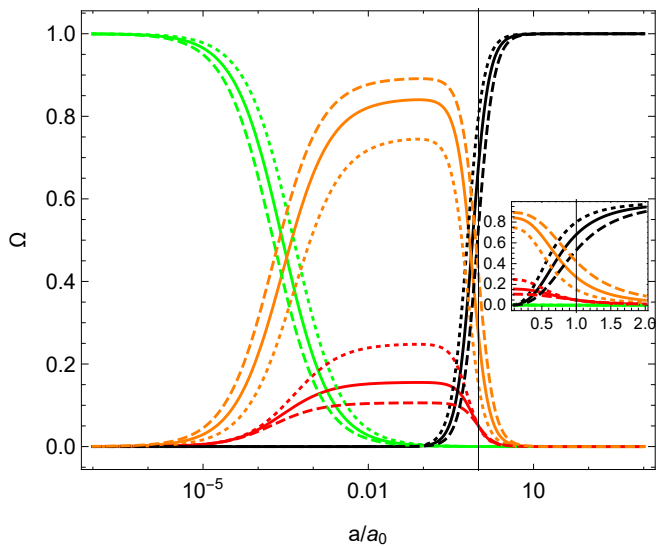}
\caption{The same as Fig.~\ref{Fig1_ev_Rastall}.
The calculated cosmic evolutions of dimensionless density parameters as functions of $a/a_0$.
The calculations are carried out for the standard $\Lambda$CDM model.
The cosmic evolution, as well as the results regarding arbitrary initial perturbations, are shown in solid, dashed, and dotted curves.
The zoomed-in plot illustrates the density parameters in the vicinity of $a/a_0=1$.
}
\lb{Fig1_ev_LCDM}
\end{figure}

To clearly illustrate the difference in the resultant cosmic evolution between the two models, we present a comparison of the calculated density parameters in Fig.~\ref{Fig1_ev_comp}.
It is found that although the density parameters of the dark energy and dark matter are identical at the present day in both models, their respective rates of change are distinct.
In the $\Lambda$CDM model, the density parameter increases rapidly at $a/a_0=1$, whereas that of the matter falls dramatically.
As a result, to reproduce their measured values at the present day, one must carefully tune the initial conditions, which, in turn, gives rise to the related coincidence problem, as illustrated in Fig.~\ref{Fig1_ev_LCDM}.
In the generalized Rastall theory, on the other hand, the evolutions of ordinary matter and dark matter are separated.
The dark matter starts to arise together with the dark energy, owing to their interaction, after the ordinary matter dominated era.
Moreover, both the dark energy and dark matter begin to saturate at the present day. Therefore their values do not sensitively depend on the initial conditions.

\begin{figure}[tbp]
\centering
\includegraphics[width=1\columnwidth]{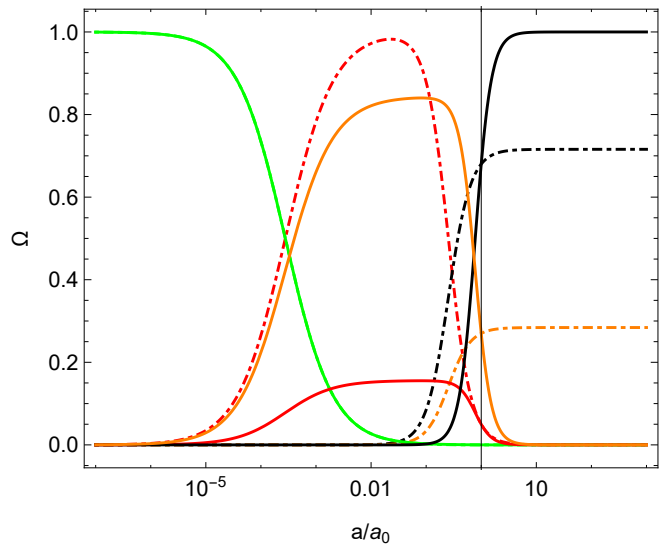}
\caption{An comparison of the calculated cosmic evolutions of the density parameters of dark energy between the two models.
The results of the standard $\Lambda$CDM model are shown in solid curves, and those of the generalized Rastall theory are indicated by dash-dotted curves.
}
\lb{Fig1_ev_comp}
\end{figure}

In Figs.~\ref{Fig2_q_Rastall}-\ref{Fig2_q_comp}, one shows the resultant deceleration parameters for different initial conditions as functions of redshift in both models.
Again, it is found that the deceleration parameter eventually approaches a given value, independent of specific initial conditions.
Regarding both models, the values of $q$ are identical at $a/a_0=0$, and the general trend is also found to be similar.
However, for the Rastall gravity, one observes that $q$ has begun to converge at $a/a_0=0$.
This is different from the case of the $\Lambda$CDM model where, again, at the present-day $q$ is falling rapidly.
As a result, the related value of $q$ is sensitively governed by the specific initial conditions. 

\begin{figure}[tbp]
\centering
\includegraphics[width=1\columnwidth]{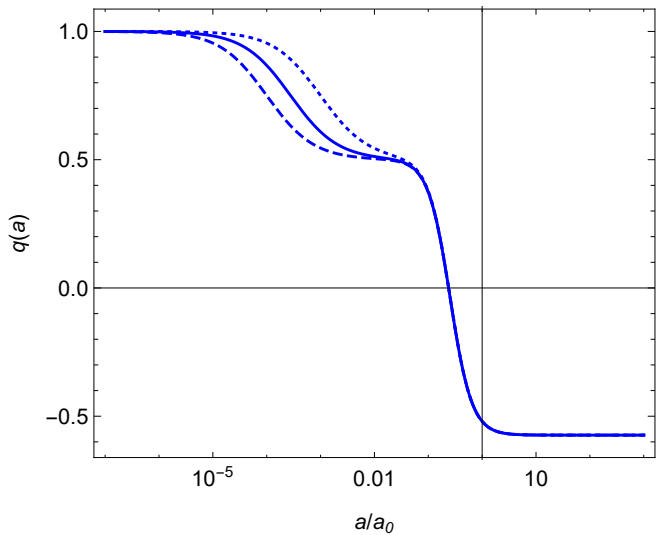}
\caption{
The calcualted deceleration parameter $q$ as a function of $a/a_0$.
The present day $a/a_0=1$ is indicated by a vertical black solid line.
The calculations are carried out for different parameters in generalized Rastall gravity.
The cosmic evolution evaluated by using the specific initial conditions which reproduces the measurements is presented by solid curves.
Those obtained by using different perturbed initial conditions are indicated by dashed and dotted curves.
}
\lb{Fig2_q_Rastall}
\end{figure}

\begin{figure}[tbp]
\centering
\includegraphics[width=1\columnwidth]{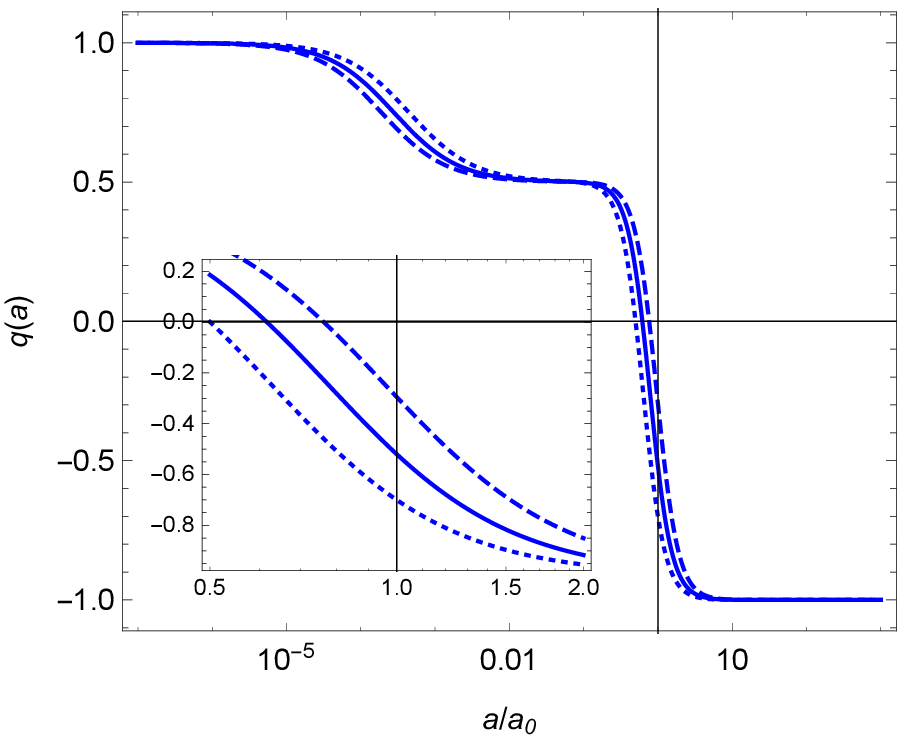}
\caption{The same as Fig.~\ref{Fig2_q_Rastall}.
The calcualted deceleration parameter $q$ as a function of $a/a_0$.
The calculations are carried out for the standard $\Lambda$CDM model.
The cosmic evolution, as well as the results regarding arbitrary initial perturbations, are shown in solid, dashed, and dotted curves.
The zoomed-in plot illustrates the deceleration parameters in the vicinity of $a/a_0=1$.
}
\lb{Fig2_q_LCDM}
\end{figure}

\begin{figure}[tbp]
\centering
\includegraphics[width=1\columnwidth]{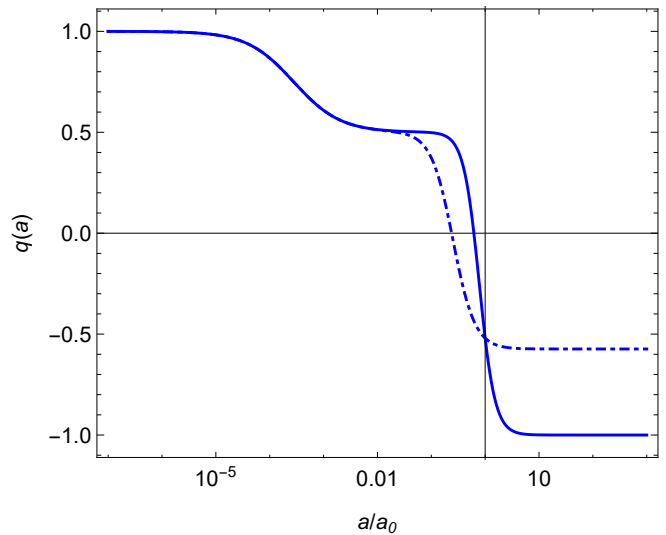}
\caption{An comparison of the calculated deceleration parameter $q$ between the two models.
The results of the standard $\Lambda$CDM model are shown in solid curves, and those of the generalized Rastall theory are indicated by dash-dotted curves.
}
\lb{Fig2_q_comp}
\end{figure}

The above properties regarding the generalized Rastall theory can be shown more transparently as one focuses on the deviations from the specific solution discussed at the beginning of the section.
The corresponding results are presented in Fig.~\ref{Fig4_attractor_Rastall} where one studies the discrepancies in cosmic evolutions by arbitrarily perturbing the initial conditions.
To illustrate, we have chosen to show the differences in the density parameter of dark energy $\Omega_{\mathrm{de}}$ and the deceleration parameter $q$.  
It is observed that the solution displays the properties of a dynamic attractor, which is reminiscent of quintessence and k-essence models.
In other words, it is found that the deviations in evolution regarding different initial conditions all converge to the origin.
Therefore, they are insensibility to the initial conditions in the present approach.

\begin{figure}[tbp]
\centering
\includegraphics[width=1\columnwidth]{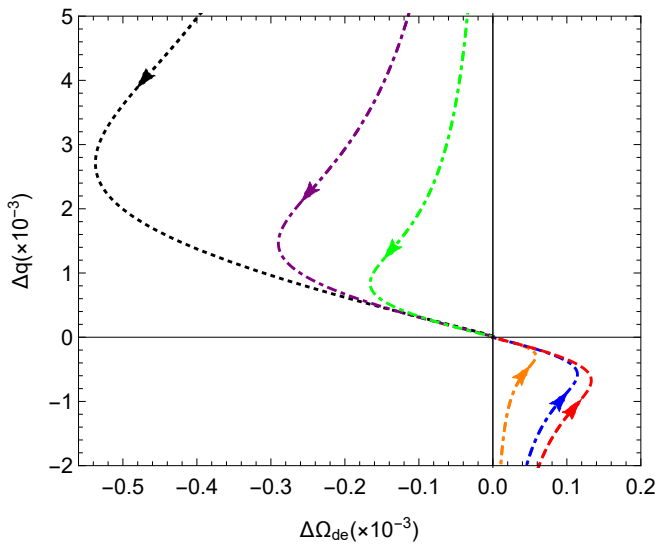}
\caption{
The dynamic attractor solution in the generalized Rastall gravity.
The results show the deviations from the specific solution governed by a specific choice of initial conditions.
The calculations are carried out for the differences in the density parameter of dark energy $\Omega_{\mathrm{de}}$ ($x$-axis) and those in deceleration parameter $q$ ($y$-axis). 
Each individual curve is obtained by evaluating the cosmic evolution with an arbitrary initial condition.
The red dashed curve and black dotted curve correspond to the same perturbations investigated in Figs.~\ref{Fig1_ev_Rastall} and \ref{Fig2_q_Rastall}.
The calculations are carried out for generalized Rastall gravity by using the parameters given in the text.}
\lb{Fig4_attractor_Rastall}
\end{figure}

Last but not least, in Fig.~\ref{Fig3_violation_Rastall}, we show the amount of deviation from a conserved SET, which is the 0-component of the r.h.s. of Eq.~\eqref{RastallEquation1}, as a function of $a/a_0$, for the generalized Rastall theory.
As discussed above, for the present model, the amount of violation is related to the dynamical evolution of the dark energy.
As shown in Fig.~\ref{Fig3_violation_Rastall}, the deviation is time-dependent.
Its magnitude becomes more significant when the dark energy evolves more rapidly, and the peak is found to locate at approximately $a/a_0 \sim 0.3$.
On the other hand, the SET is mostly conserved in the infinite past and future.

\begin{figure}[tbp]
\centering
\includegraphics[width=1\columnwidth]{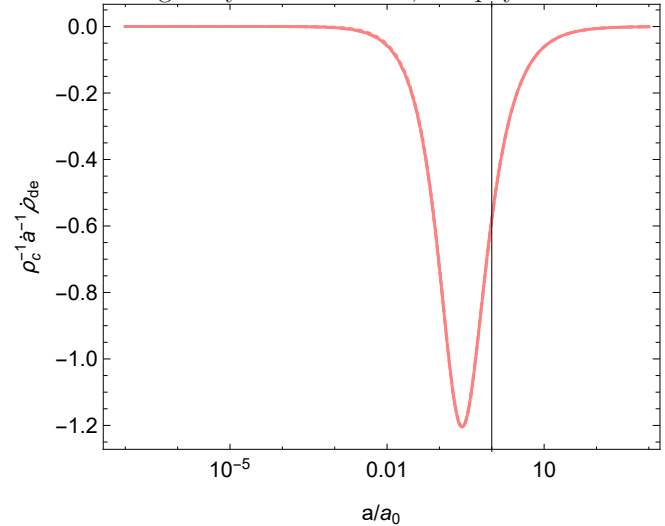}
\caption{
The amount of deviation from a conserved SET, shown as a function of $a/a_0$.
The calculations are carried out for generalized Rastall gravity by using the model parameters described in the text.}
\lb{Fig3_violation_Rastall}
\end{figure}

\section{Discussions and concluding remarks}
\renewcommand{\theequation}{5.\arabic{equation}} \setcounter{equation}{0}

Owing to the fact that one has to discard one of the solutions of Eq.~\eqref{H2}, which introduces a vanishing factor on both sides of Eq.~\eqref{H1}, the numerator of the equation is chosen as a second-order polynomial. 
In fact, Eq.~\eqref{H1} only contains $\Theta$ as an unknown scalar function, which is identified with a dynamical cosmological constant.
For this reason, it is actually a rather economical choice of {\it ansatz} in the present model.

In comparison to the standard $\Lambda$CDM model, effectively, the proposed scheme only contains one additional variable, $\Theta$.
The latter is described by the assumed equation of motion Eq.~\eqref{CosmologyEquations5}.
In this context, it is a minimal scheme necessarily to describe the dynamical evolution of dark energy.
In comparison to other recent studies~\cite{agr-rastall-cosmo-10, agr-rastall-cosmo-11} about cosmic evolution in Rastall theory, the present approach introduces a unified scheme to deal with different matter contents of the Universe.
In other words, by solving a closed system of equations, different eras of cosmic evolution are derived naturally.
Moreover, we argue that our model possesses a dynamic attractor solution, which provides a possible explanation for the coincidence problem.

It is also interesting to mention that the non-conserved SET can be treated in terms of a generalized version of the two measure theories~\cite{agr-modified-gravity-two-measure-01,agr-modified-gravity-two-measure-02}.
In this case, the dynamics can be derived from an action which consists of two measures.
In particular, the latter involves a scalar density in the place of the usual factor of the Jacobian $\sqrt{-g}$.
The theory is recently generalized in order to accommodate the fact the SET is not conserved as one considers the diffusive process in the relativistic Fokker-Plank equation~\cite{agr-modified-gravity-FP-01}.
There, the divergence of the SET is shown to be related to the conserved particle flow. 
This can be achieved by replacing the dynamic spacetime four-vector in the original theory by the gradient of a scalar field~\cite{agr-modified-gravity-two-measure-03,agr-modified-gravity-two-measure-04}.
The resultant theory gives rise to a unified description of the interacting dark energy and dark matter.
It is, therefore, intriguing to compare the above approach against the Lagrangian formalism of Rastall theory.

To summarize, the present study involves an attempt to propose a scheme for cosmic evolution in a generalized Rastall gravity.
In our model, the physical content of the dark energy is attributed to the non-conserved sector of the SET.
The resultant cosmic evolution is naturally found to consists of three stages, namely, radiation dominated, ordinary matter dominated, as well as dark energy and dark matter dominated eras.
Also, for the present model, it is shown that the eventual fate of the Universe is largely insensitive to the initial conditions, and the cosmic coincidence problem is therefore averted.
Furthermore, we show that the amount of violation is found to be more significant when the dark energy evolves dynamically.

\begin{acknowledgments}
We gratefully acknowledge the financial support from
Funda\c{c}\~ao de Amparo \`a Pesquisa do Estado de S\~ao Paulo (FAPESP),
Funda\c{c}\~ao de Amparo \`a Pesquisa do Estado do Rio de Janeiro (FAPERJ),
Conselho Nacional de Desenvolvimento Cient\'{\i}fico e Tecnol\'ogico (CNPq),
Coordena\c{c}\~ao de Aperfei\c{c}oamento de Pessoal de N\'ivel Superior (CAPES),
and National Natural Science Foundation of China (NNSFC) under contract No. 11805166.
\end{acknowledgments}

\bibliographystyle{h-physrev}
\bibliography{references_qian}

\begin{thebibliography}{10}

\bibitem{agr-rastall-01}
P.~Rastall,
\newblock Phys. Rev. {\bf D6}, 3357 (1972).

\bibitem{agr-rastall-02}
Y.~Heydarzade and F.~Darabi,
\newblock Phys. Lett. {\bf B771}, 365 (2017), arXiv:1702.07766.

\bibitem{agr-rastall-03}
J.~P. Morais~Graca and I.~P. Lobo,
\newblock Eur. Phys. J. {\bf C78}, 101 (2018), arXiv:1711.08714.

\bibitem{agr-rastall-04}
K.~A. Bronnikov, J.~C. Fabris, O.~F. Piattella, and E.~C. Santos,
\newblock Gen. Rel. Grav. {\bf 48}, 162 (2016), arXiv:1606.06242.

\bibitem{agr-rastall-05}
Y.~Heydarzade, H.~Moradpour, and F.~Darabi,
\newblock Can. J. Phys. {\bf 95}, 1253 (2017), arXiv:1610.03881.

\bibitem{agr-rastall-06}
E.~Spallucci and A.~Smailagic,
\newblock Int. J. Mod. Phys. {\bf D27}, 1850003 (2017), arXiv:1709.05795.

\bibitem{agr-rastall-09}
R.~Kumar and S.~G. Ghosh,
\newblock Eur. Phys. J. {\bf C78}, 750 (2018), arXiv:1711.08256.

\bibitem{agr-rastall-11}
Z.~Xu, X.~Hou, X.~Gong, and J.~Wang,
\newblock Eur. Phys. J. {\bf C78}, 513 (2018), arXiv:1711.04542.

\bibitem{agr-rastall-15}
M.~Sadeghi,
\newblock (2018), arXiv:1809.08698.

\bibitem{agr-rastall-16}
I.~P. Lobo, H.~Moradpour, J.~P. Morais~Graça, and I.~G. Salako,
\newblock Int. J. Mod. Phys. {\bf D27}, 1850069 (2018), arXiv:1710.04612.

\bibitem{agr-rastall-21}
K.~Lin, Y.~Liu, and W.-L. Qian,
\newblock Gen. Rel. Grav. {\bf 51}, 62 (2019), arXiv:1809.10075.

\bibitem{agr-rastall-22}
K.~Lin and W.-L. Qian,
\newblock Chin. Phys. {\bf C43}, 083106 (2019), arXiv:1812.10100.

\bibitem{agr-rastall-cosmo-01}
A.~S. Al-Rawaf and M.~O. Taha,
\newblock Phys. Lett. {\bf B366}, 69 (1996).

\bibitem{agr-rastall-cosmo-02}
A.~S. Al-Rawaf and M.~O. Taha,
\newblock Gen. Rel. Grav. {\bf 28}, 935 (1996).

\bibitem{agr-rastall-cosmo-03}
C.~E.~M. Batista, M.~H. Daouda, J.~C. Fabris, O.~F. Piattella, and D.~C.
  Rodrigues,
\newblock Phys. Rev. {\bf D85}, 084008 (2012), arXiv:1112.4141.

\bibitem{agr-rastall-cosmo-04}
J.~C. Fabris, O.~F. Piattella, D.~C. Rodrigues, C.~E.~M. Batista, and M.~H.
  Daouda,
\newblock Int. J. Mod. Phys. Conf. Ser. {\bf 18}, 67 (2012), arXiv:1205.1198.

\bibitem{agr-rastall-cosmo-05}
K.~A. Bronnikov, J.~C. Fabris, O.~F. Piattella, D.~C. Rodrigues, and E.~C.
  Santos,
\newblock Eur. Phys. J. {\bf C77}, 409 (2017), arXiv:1701.06662.

\bibitem{agr-rastall-cosmo-06}
F.~Darabi, K.~Atazadeh, and Y.~Heydarzade,
\newblock Eur. Phys. J. Plus {\bf 133}, 249 (2018), arXiv:1710.10429.

\bibitem{agr-rastall-cosmo-07}
F.-F. Yuan and P.~Huang,
\newblock Class. Quant. Grav. {\bf 34}, 077001 (2017), arXiv:1607.04383.

\bibitem{agr-rastall-cosmo-08}
J.~C. Fabris, M.~H. Daouda, and O.~F. Piattella,
\newblock Phys. Lett. {\bf B711}, 232 (2012), arXiv:1109.2096.

\bibitem{agr-rastall-cosmo-09}
C.~E.~M. Batista, J.~C. Fabris, O.~F. Piattella, and A.~M. Velasquez-Toribio,
\newblock Eur. Phys. J. {\bf C73}, 2425 (2013), arXiv:1208.6327.

\bibitem{agr-rastall-cosmo-10}
H.~Moradpour, Y.~Heydarzade, F.~Darabi, and I.~G. Salako,
\newblock Eur. Phys. J. {\bf C77}, 259 (2017), arXiv:1704.02458.

\bibitem{agr-cosmo-review-01}
S.~Weinberg,
\newblock Rev. Mod. Phys. {\bf 61}, 1 (1989),
\newblock [,569(1988)].

\bibitem{agr-cosmo-review-02}
T.~Padmanabhan,
\newblock Phys. Rept. {\bf 380}, 235 (2003), arXiv:hep-th/0212290.

\bibitem{agr-cosmo-review-03}
E.~J. Copeland, M.~Sami, and S.~Tsujikawa,
\newblock Int. J. Mod. Phys. {\bf D15}, 1753 (2006), arXiv:hep-th/0603057.

\bibitem{agr-cosmo-review-04}
K.~Bamba, S.~Capozziello, S.~Nojiri, and S.~D. Odintsov,
\newblock Astrophys. Space Sci. {\bf 342}, 155 (2012), arXiv:1205.3421.

\bibitem{agr-cosmo-review-05}
M.~Li, X.-D. Li, S.~Wang, and Y.~Wang,
\newblock Commun. Theor. Phys. {\bf 56}, 525 (2011), arXiv:1103.5870.

\bibitem{agr-cosmo-de-quintessence-01}
R.~R. Caldwell, R.~Dave, and P.~J. Steinhardt,
\newblock Phys. Rev. Lett. {\bf 80}, 1582 (1998), arXiv:astro-ph/9708069.

\bibitem{agr-cosmo-de-tachyon-01}
A.~Sen,
\newblock JHEP {\bf 07}, 065 (2002), arXiv:hep-th/0203265.

\bibitem{agr-cosmo-de-k-essence-01}
C.~Armendariz-Picon, V.~F. Mukhanov, and P.~J. Steinhardt,
\newblock Phys. Rev. Lett. {\bf 85}, 4438 (2000), arXiv:astro-ph/0004134.

\bibitem{agr-cosmo-de-phantom-01}
S.~Nojiri and S.~D. Odintsov,
\newblock Phys. Lett. {\bf B562}, 147 (2003), arXiv:hep-th/0303117.

\bibitem{agr-cosmo-de-chaplygin-01}
A.~{\relax Yu}. Kamenshchik, U.~Moschella, and V.~Pasquier,
\newblock Phys. Lett. {\bf B511}, 265 (2001), arXiv:gr-qc/0103004.

\bibitem{agr-cosmo-de-holo-01}
A.~G. Cohen, D.~B. Kaplan, and A.~E. Nelson,
\newblock Phys. Rev. Lett. {\bf 82}, 4971 (1999), arXiv:hep-th/9803132.

\bibitem{agr-cosmo-de-holo-02}
M.~Li,
\newblock Phys. Lett. {\bf B603}, 1 (2004), arXiv:hep-th/0403127.

\bibitem{agr-cosmo-de-holo-04}
B.~Wang, Y.-g. Gong, and E.~Abdalla,
\newblock Phys. Lett. {\bf B624}, 141 (2005), arXiv:hep-th/0506069.

\bibitem{agr-cosmo-de-agegraphic-01}
R.-G. Cai,
\newblock Phys. Lett. {\bf B657}, 228 (2007), arXiv:0707.4049.

\bibitem{agr-cosmo-de-agegraphic-02}
H.~Wei and R.-G. Cai,
\newblock Phys. Lett. {\bf B660}, 113 (2008), arXiv:0708.0884.

\bibitem{agr-modified-gravity-fR-review-01}
A.~De~Felice and S.~Tsujikawa,
\newblock Living Rev. Rel. {\bf 13}, 3 (2010), arXiv:1002.4928.

\bibitem{agr-modified-gravity-fT-review-01}
F.~W. Hehl, J.~D. McCrea, E.~W. Mielke, and Y.~Ne'eman,
\newblock Phys. Rept. {\bf 258}, 1 (1995), arXiv:gr-qc/9402012.

\bibitem{agr-modified-gravity-fRT-01}
T.~Harko, F.~S.~N. Lobo, S.~Nojiri, and S.~D. Odintsov,
\newblock Phys. Rev. {\bf D84}, 024020 (2011), arXiv:1104.2669.

\bibitem{agr-modified-gravity-BD-01}
C.~Brans and R.~H. Dicke,
\newblock Phys. Rev. {\bf 124}, 925 (1961),
\newblock [,142(1961)].

\bibitem{agr-modified-gravity-GB-01}
S.~Nojiri and S.~D. Odintsov,
\newblock Phys. Lett. {\bf B631}, 1 (2005), arXiv:hep-th/0508049.

\bibitem{agr-modified-gravity-Lovelock-01}
D.~Lovelock,
\newblock J. Math. Phys. {\bf 12}, 498 (1971).

\bibitem{agr-modified-gravity-horava-01}
P.~Horava,
\newblock Phys. Rev. {\bf D79}, 084008 (2009), arXiv:0901.3775.

\bibitem{agr-modified-gravity-horava-02}
P.~Horava,
\newblock Phys. Rev. Lett. {\bf 102}, 161301 (2009), arXiv:0902.3657.

\bibitem{agr-modified-gravity-horava-lw-01}
K.~Lin, A.~Wang, Q.~Wu, and T.~Zhu,
\newblock Phys. Rev. {\bf D84}, 044051 (2011), arXiv:1106.1486.

\bibitem{agr-modified-gravity-horava-lw-05}
K.~Lin, S.~Mukohyama, A.~Wang, and T.~Zhu,
\newblock Phys. Rev. {\bf D89}, 084022 (2014), arXiv:1310.6666.

\bibitem{agr-rastall-23}
W.~A.~G. De~Moraes and A.~F. Santos,
\newblock Gen. Rel. Grav. {\bf 51}, 167 (2019), arXiv:1912.06471.

\bibitem{agr-cosmo-bb-04}
G.~W. Gibbons and S.~W. Hawking,
\newblock Phys. Rev. {\bf D15}, 2738 (1977).

\bibitem{agr-cosmo-bb-05}
L.~Parker,
\newblock Phys. Rev. {\bf D3}, 346 (1971),
\newblock [Erratum: Phys. Rev.D3,2546(1971)].

\bibitem{agr-cosmo-bb-06}
L.~H. Ford,
\newblock Phys. Rev. {\bf D35}, 2955 (1987).

\bibitem{book-qft-Bertlemann}
R.~Bertlemann.,
\newblock {\em {Anomalies in quantum field theory}} (Oxford university press,
  1996).

\bibitem{book-qft-curved-space-Birrell-Davies}
N.~D. Birrell and P.~C.~W. Davies,
\newblock {\em {Quantum Fields in Curved Space}} (Cambridge University Press,
  1982).

\bibitem{agr-modified-gravity-FP-01}
S.~Calogero,
\newblock JCAP {\bf 11}, 016 (2011), arXiv:1107.4973.

\bibitem{agr-rastall-cosmo-11}
D.~Das, S.~Dutta, and S.~Chakraborty,
\newblock Eur. Phys. J. {\bf C78}, 810 (2018), arXiv:1810.11260.

\bibitem{agr-rastall-18}
F.~Darabi, H.~Moradpour, I.~Licata, Y.~Heydarzade, and C.~Corda,
\newblock Eur. Phys. J. {\bf C78}, 25 (2018), arXiv:1712.09307.

\bibitem{agr-modified-gravity-two-measure-03}
D.~Benisty and E.~Guendelman,
\newblock Eur. Phys. J. C {\bf 77}, 396 (2017), arXiv:1701.08667.

\bibitem{agr-modified-gravity-two-measure-04}
D.~Benisty, E.~Guendelman, and Z.~Haba,
\newblock Phys. Rev. D {\bf 99}, 123521 (2019), arXiv:1812.06151,
\newblock [Erratum: Phys.Rev.D 101, 049901 (2020)].

\bibitem{agr-cosmo-data-planck-01}
Planck, P.~A.~R. Ade {\em et~al.},
\newblock Astron. Astrophys. {\bf 571}, A1 (2014), arXiv:1303.5062.

\bibitem{book-cosmology-Gregory-Zeilik}
S.~A. Gregory and M.~Zeilik,
\newblock {\em Introductory Astronomy and Astrophysics}Saunders Golden Sunburst
  Series, 4 ed. (Cengage Learning, 1997).

\bibitem{agr-cosmo-de-holo-03}
S.~D.~H. Hsu,
\newblock Phys. Lett. {\bf B594}, 13 (2004), arXiv:hep-th/0403052.

\bibitem{agr-modified-gravity-two-measure-01}
E.~Guendelman and A.~Kaganovich,
\newblock Phys. Rev. D {\bf 53}, 7020 (1996), arXiv:gr-qc/9605026.

\bibitem{agr-modified-gravity-two-measure-02}
E.~Guendelman,
\newblock Mod. Phys. Lett. A {\bf 14}, 1043 (1999), arXiv:gr-qc/9901017.

\end{thebibliography}
\end{document}